\documentstyle[aps,eqsecnum,epsf]{revtex}
\begin{document}
\input epsf.tex
\title{Evaluation of Multiloop Diagrams via Lightcone Integration}
\author{Y.J. Feng\cite{YJF} and C. S. Lam\cite{CSL}}
\address{Department of Physics, McGill University,
3600 University St., Montreal, P.Q., Canada H3A 2T8}
\maketitle

\begin{abstract}
We present a systematic method to determine the dominant regions of
internal momenta contributing to any two-body
high-energy near-forward scattering diagram. Such a knowledge is used
to evaluate leading high-energy dependences of loop diagrams.
It also gives a good idea where dominant multiparticle cross sections occur.

\end{abstract}

\section{Introduction}
It is difficult to compute high-energy ($\sqrt{s}$)
 scattering amplitudes at small momentum transfers ($\sqrt{-t}$), 
even assuming the coupling constant $g^2$
to be small. 
 This is so because each loop of a Feynman diagram 
is capable of producing a $\ln s$ factor, thus
changing the effective expansion parameter from $g^2$ to $g^2\ln s$. 
Even though the former may be small, the latter
can become quite sizable at high energies, necessitating 
diagrams of high orders to be 
included. Such is for example the case
when total cross section is computed in the framework of QCD \cite{lam}.

Usually such a daunting task of computed diagrams of
many loops  may be contemplated only
in the leading-log approximation (LLA), though there are exceptions, 
especially for sets of diagrams with regular structures
\cite{fadlip}. In LLA ,
only terms of the highest power of $\ln s$ are kept at each
perturbative order,
 but even so the computation is far from being simple.
For low-order diagrams, or diagrams with highly regular structures,
the computation has been carried out and the results are well known
 \cite{POLK,cw,bfkl,bartels}. However, for complicated diagrams, a 
systematic procedure to find even the leading-log contribution
seems to be lacking. We shall discuss a method in the present
 paper. The available QCD result, via the exchange
of the BFKL Pomeron \cite{bfkl}, violates the Froissart bound and
needs to be improved \cite{reggeon}.
Other diagrams must be included to restore unitarity so it would be
useful to have a way to find out how the other diagrams behave at
high energies. This can be achieved if 
the regions of internal momenta dominating the Feynman amplitude
can be located, for then 
one simply integrates around
them to obtain the LLA result.  

For quark-quark scattering via the exchange of gluon ladders, 
the dominant region is known to be the
multi-Regge region \cite{cw,bfkl}, where gluons
produced in the intermediate states are strongly ordered
in rapidity, and the gluons being exchanged are dominantly
spacelike. What we would like to discuss in the present paper
 is a general way to find such dominant
regions for any diagram, and its associated high-energy
dependence in LLA. We shall carry out the study for Feynman diagrams
and for nonabelian cut diagrams \cite{FHL,FL,LL}, both because they are
more general, and because there is already a considerable body
of literature on the dispersion theoretic techniques \cite{bfkl,bartels}.

Such calculations of elastic amplitudes, besides giving the
energy-dependence of total cross-sections via the optical theorem
\cite{lam},
also tell us the kinematical regions where the dominant inelastic
cross-sections come from, for via unitarity these are intimately
related to the dominant internal momenta of the elastic amplitude.
This knowledge would be of direct phenomenological interest as well.

The methods developed in this paper should also be useful in the 
study of two-dimensional effective QCD Lagrangians at high-energies 
\cite{verlinde,lipatov}. A prerequisite needed to arrive at a reliable
effective Lagrangian is to know which are the heavy modes that
can be discarded, and which of them must be integrated out to 
yield a new vertex in the effective Lagrangian. In perturbative language
this is equivalent to finding the important regions of
internal momenta around which to integrate. All others may simply
be discarded.

In the rest of this section we shall describe what our method
is based on, and provide a brief summary of the results.

At high energies it is convenient
 to use lightcone coordinates, $k_\pm=k^0\pm k^3$.
The components
of a four-vector $k^\mu$ can then be written as
$(k_+,k_-,k_\perp)$, and the loop integration 
expressed as $d^4k=dk_+dk_-d^2k_\perp/2$. 
In the centre-of-mass system, the momenta of the two
incoming particles, with masses neglected, can be taken
to be $p_2=(\sqrt{s},0,0)$ and $p_1=(0,\sqrt{s},0)$.
The momentum transfer $\sqrt{-t}$ as well as all 
other transverse momenta $k_\perp$ are taken to be of
order 1 as $s\to \infty$, so it is only the dominant
regions in $k_+$ and $k_-$ for every loop momentum $k$
that have to be determined.

These regions are determined in the following 
way. First, observe that the inverse of the internal propagators
are bilinear in the `+' and the `$-$' components
of their line momenta, so the propagators give rise to simple
poles in
the `+' (or the `$-$') momenta which enable
integrations in those variables to be carried
out exactly by residue
calculus \cite{cw}. Once this is done the locations of
the `+' momenta are determined by the locations
of the contributing poles and the `$-$' momenta.
The `$-$' momenta are then fixed to be
in the regions 
yielding the leading-log contributions to the 
amplitude.

We shall be able to do this both for  Feynman diagrams
and  `nonabelian cut diagrams' \cite{FHL,FL,LL}.
Feynman diagrams are fundamental, but they 
often have the undesirable property that
the LLA contributions of individual diagrams get cancelled
in the sum \cite{cw}. 
To the extent that the usual technology only
allows LLA to be computed, this cancellation is disastrous
because it leaves no viable means to compute the leading high-energy
 behaviour of the sum. 
Nonabelian cut diagrams are designed to
combat this problem. The cancellation
is actually a result of the destructive interference
between the virtual gluons being exchanged. The nonabelian cut diagrams
allow the destructive interferences to take place before
high energy approximations are taken. In this way the
LLA contribution to the nonabelian cut diagrams
will reflect directly the leading contributions to the
sum. No further cancellation will occur.

In Sec.~2 we will review the {\it flow diagram} method
 of
Cheng and Wu \cite{cw} for carrying out the `+' component
integrations
by residue calculus. This method is very effective
in locating the poles and the dominant integration regions
for relatively simple diagrams. In complicated diagrams one
encounters the  problem of
{\it flow reversal} which will be discussed in Sec.~3. This
problem prevents a simple reading of the contributing poles
directly from the  flow diagrams. Nevertheless the locations of
these poles can still be computed, but the complexity of 
computation grows quite fast with the number of loops of the
diagram. This difficulty is then overcome by a `path' method 
to be discussed in Sec.~4.
With this mehtod the contributing poles can be located and the `+'
momenta determined. What remains is to find the dominant `$-$'
momenta that give rise to the LLA contribution. The recipe for 
doing so will be discussed in Sec.~5.
 Finally, in Sec.~6,
a number of examples are given to illustrate the procedure.

\section{Flow Diagrams}

Consider a diagram 
with $n$ internal lines and $\ell$ loops, whose line and loop
momenta are denoted by $q_i\ (1\le i\le n)$ and $k_b\ (1\le b\le\ell)$
respectively.  In lightcone coordinates, the denominator of the propagator
for a line with momentum $q$ is
$d(q)=(q^2-m^2+i\epsilon)=(q_+q_--q_\perp^2-m^2+i\epsilon)
\equiv (q_+q_--a+i\epsilon)$.
These $n$ propagators collectively define a set of poles for the
integration variables $k_{b+}$, thus enabling
these integrations to be performed with the help of
residue calculus. To carry out this program we must identify,
for each $k_{b+}$, which are the poles in the upper-half plane and
which are the poles in the lower-half plane, for only the poles in 
one half-plane
will be picked up by a  contour integration. Their locations in turn depends
on the sign of $q_{i-}$, the choice of loop momenta, 
as well as the order the $k_{b+}$ integrations are carried out. 
With so many variables the problem is very complex indeed.
{\it Flow diagram} was invented \cite{cw}
to keep track of things
and to determine the location of poles. We shall review
its essence \cite{cw} in this section, and point out in the next section
some of the complications hitherto overlooked. This complication makes
it complicated to apply it to multiloop diagrams.
In Sec.~4 we shall propose a 
{\it `path method'} to bypass these complications, 
and enables the
evaluation of the `+' integration to be carried out
in a simple manner.

A flow diagram is a Feynman diagram 
(or a nonabelian cut diagram) 
with arrows attached to each of its internal lines to
indicate the direction
of $q_{i-}$. Since the signs of the
 $q_-$'s vary over the
integration region, generally more than one
 flow diagram is present for each
Feynman or nonabelian cut diagram. Nevertheless, for a diagram
with $n$ internal lines, there are far fewer than $2^n$ flow diagrams
that one might otherwise expect,
for two reasons. First, momentum conservation forbids the arrows 
from a common vertex to point all inwards or all outwards. 
Secondly, for reasons to be explained below, one can reject
flow diagrams in which arrows around any
 closed loop all
point in the same (clockwise or  counter-clockwise) direction.
With these two requirements, 
it is easy to see that the 1-loop box diagram
has only one flow diagram, rather than $2^4=16$.

In a flow diagram the signs of $q_{i-}$ along the arrows are 
all positive, by definition. This allows the positions
of the poles be located and the `$+$' integrations
to be carried out, once the independent loops and their order of integrations are chosen. We shall now proceed to see how this is 
accomplished for the first integration, say
$k_{1+}$.

$k_{1+}$ flows through the lines of this first loop
either in a clockwise or
a counter-clockwise direction. Its coefficient in $d(q_i)$ is
$\pm q_{i-}$, depending on whether this direction is the same as
the arrow or opposite. The pole of $1/d(q_i)$ in $k_{1+}$
 has an imaginary part $\mp i\epsilon/q_{i-}$, with all
$q_{i-}>0$ by definition. Hence the lines with arrows pointing
one way (clockwise or counter-clockwise) have poles all in
one half-plane, and those with arrows pointing the opposite way have
poles in the other half-plane. Which is which does not matter because
we can always define the loop momentum by reversing its sign. 

It is now easy to understand the assertion made earlier in the section,
that flow diagrams containing a closed loop with flow arrows all pointing
in the same direction may be rejected. Taking this loop as the first
loop of integration, this would imply all poles to be in the same
half-plane. By closing the integration contour in the other
half-plane, we get a zero integral so such a flow diagram can be ignored.

Sometimes pole locations for {\it subsequent integrations} can be
located in the same way, {\it i.e.,} by the direction of arrows in
the flow diagram.
In fact the explicit examples shown
in Ref.~\cite{cw} all seem to be of this type.

However, it is not guaranteed that pole locations for subsequent
integrations can be located this way, as we shall now see. This is
the complication mentioned in the section.

To make it easier to describe things later on, we shall call two
momenta pointing in the same (opposite) direction around a loop
to be {\it parallel (anti-parallel)} in that loop.

\section{Flow Reversal}

Suppose there are $n_1$ poles picked up by the $k_{1+}$ integration,
each contributing to a term in the integral. 
As a result of the integration,
 $k_{1+}$ acquires an imaginary part
$\mp i\epsilon/q_{i-}$ from the $i$th pole.
The sign is $-/+$ if  $k_{1+}$ and
$q_{i-}$ are parallel/anti-parallel. This imaginary part
in turn imparts an imaginary part on every $q_{j+}$ of the first loop,
which is why the location of poles for the second and subsequent 
integrations may be altered. For simplicity, we shall assume from now 
on that $\epsilon$ is finite and positive, and has a common value in all 
the propagators.

This imaginary part of $k_{1+}$ affects the location of poles 
in subseqent integrations only for lines $j$ lying in loop 1. 
In that case,  the 
imaginary part of $d(q_j)$ is 
changed from $i\epsilon$ to  $i\epsilon(\mp q_{j-}/q_{i-}+1)$, with
sign $-/+$ when lines $j$ and $i$ are parallel/anti-parallel in loop 1.
Unless the sign is $-$ and
$q_j>q_i$, the imaginary part of $d(q_j)$
remains positive and the location of pole $j$ 
in subsequent integrations is once again determined solely
by the direction of its arrow around the integration loop,
{\it viz.,} it can be determined directly from the flow
diagram.
However, if lines $j$ and $i$ 
are parallel in the first loop, and that $q_j>q_i$,
then the sign of the imaginary part of $d(q_j)$ becomes
negative, and the pole location (upper or lower plane) will now be opposite
to naive expectations from the flow diagram.
This situation can still be accommodated into the
flow diagram if we simply reverse the arrow of this line by hand.
This is {\it flow reversal}. 

To summarize, here is how poles for the `+'
integrations are computed for a given flow diagram, assuming a set of
independent loops and a given order of
 $k_{b+}$-integrations have been chosen.

For the first loop, use the {\it naive rule}
to read it off the flow diagram. This means that
lines of this loop with arrows pointing in the same direction
have their poles in the same half plane.

Assuming now $k_{b+}$-integrations have been carried out for
$b=1,2,\cdots,c$. We shall now proceed to do the $(c+1)$th
integration for the term resulting from picking up
 poles located at line $i_b$ for
the $b$th loop, $b=1,2,\cdots,c$.

First note that whatever loop $(c+1)$ is, it should not contain 
any of the lines $i_1,i_2,\cdots,i_c$. This is because the
`+' momenta of these lines have been determined by previous integrations
so they cannot be fixed again by the $(c+1)$th integration.

The naive rule can be used for lines $j$ in loop $(c+1)$
if, (i) it is not in any one of the previous loops, $1,2,\cdots,c$,
(ii) it is in a previous loop $b$ but $j$ is anti-parallel to $i_b$
in that loop,
or (iii) $j$ is parallel to $i_b$ around loop
$b$ but $q_{j-}<q_{i_b-}$. In the remaining case, when
$j$ and $i_b$ are parallel in loop
$b$ but $q_{j-}>q_{i_b-}$, we must reverse the
arrow direction of line $j$ before the naive rule is applied.

After all $\ell$ `+'-integrations are carried out, we obtain a number
of terms, each of which is specified by a set of poles $i_b$
for loop $b$. We shall call this collection of lines,
$I=(i_1i_2\cdots i_\ell)$, a {\it contributing pole}.

Let us illustrate this recipe of obtaining
 contributing poles with two explicit examples: a two-loop
diagram, and a four-loop diagram. In the process we will see how important
it is to take  flow reversals into account 
just to maintain consistency.

\subsection{A two-loop example}
Fig.~1 is one of two possible flow diagrams for a two-loop Feynman diagram;
the other has line 6 reversed.

Let $a$ denote the loop with lines (1536) and $b$ the loop with lines
(2647). The big loop with lines (153472) is the union of these
two loops and will be denoted by
$a.b$. Only two
of the three loop-momenta are independent.

\begin{figure}
\vskip -0 cm
\centerline{\epsfxsize 3 truein \epsfbox {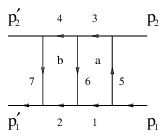}}
\nobreak
\vskip -2.5 cm\nobreak
\vskip .1 cm
\caption{A two-loop (Feynman) flow diagram.}
\end{figure}

There are three ways to start out the first loop integration,
but the final results of their integrals must be the identical,
and we must be able to pick the same 
contributing poles as well. We will illustrate here 
in detail
how the latter can be achieved, {\it iff} proper flow
reversals are taken into account.

Suppose we first integrate over loop $a$.
In this loop the arrow of line 1 and the arrows of lines 5,3,6 are
opposite, so their respective poles lie in opposite half
planes of $k_{a+}$. We shall pick line 1 to be the 
relevant pole for further discussions.
To simplify later descriptions we shall abbreviate this process
of picking pole 1 from loop $a$ simply by $a(1)$.

Now we are ready to tackle the second integration.
Since line 1 is on $a.b$ we must not choose $a.b$
to be the second loop, so we are forced to choose 
it to be $b$.
Line 6, which is in both loops $b$ and $a$, is anti-parallel
to line 1 in loop $a$, so the naive rule once again applies
to loop $b$. Lines 6 and 2 are on one half-plane, and lines 4 and
7 on the other. We shall pick 2 and 6 to be the relevant poles.
Consequently, we obtain two contributing poles, $I_1=(1,2)$ from 
$a(1)b(2)$, and $I_2=(1,6)$ from $a(1)b(6)$.

Next, let us start all over again but this time first carry
out the integration around loop $b$ to get $b(2)$ and $b(6)$.
Now since line 2 lies in $a.b$, for the term $b(2)$ the second
loop must be chosen to be $a$. 
In loop $b$ line $6$ is parallel to line $2$,
so there is a chance it might suffer a flow reversal. However,
since $q_{2-}=q_{6-}+q_{1-}>q_{6-}$, flow reversal does not
occur. Hence we have $b(2)a(1)$, so this contributing pole 
is $I_1=(1,2)$. For the term $b(6)$, since $6$ is in $a$, the second
loop must be chosen to be $a.b$. Now lines 2,4,7 are all in the
first loop $b$, but 4 and 7 will not suffer flow reversal because
they are antiparallel to 6. Line 2 is a different matter since
$q_{2-}>q_{6-}$, so it would suffer a flow reversal. With this 
reversal, all lines in $a.b$ point in the same direction, with
the sole exception of 1, so this yields $b(6)a.b(1)$, and the
contributing pole is $I_2=(1,6)$. In this way we obtain the same
set of contributing pole as before, as we should.

Finally suppose we carry out $a.b$ first, getting two terms
$a. b(1)$ and $a. b(2)$. In the first case line 1 is in $a$ so the
second loop must be $b$. Lines 4 and 7 in $b$ are antiparallel to
1 so they do not suffer from flow reversal. Line 2 is parallel to
1 and $q_{2-}>q_{1-}$ so it does suffer a flow reversal, thus 
leaving behind only line 6 of loop $b$ in one direction.
From $a.b(1)$ we therefore obtain $a.b(1)b(6)$ and the contributing
pole $I_2=(1,6)$. Now consider the term $a.b(2)$. The second loop
must now be $a$. Lines 5, 3 are antiparallel to 2 so they do not
suffer flow reversal. Line 1 is parallel to 2 but $q_{1-}<q_{2-}$,
so it does not suffer from flow reversal either. So no flow reversal
occurs at all for lines in loop $a$, and this term yields $a.b(2)a(1)$,
giving rise to the contributing pole $I_1=(1,2)$.  
The result is once again
the same as the other two calculations. If flow reversals were not
properly taken into account, the result would have been 
different and wrong.

The main lesson learned from this very
simple example is that generally detailed loop-by-loop
calculation must be performed, with proper
flow reversals taken
into account, in order to obtain the correct
locations of the contributing poles. Also, the amount of calculations
needed to determine the contributing poles may depend critically on
the independent loops chosen and the order of 
 integrations performed.

\subsection{A four-loop example}
The task of obtaining the contributing poles becomes
more arduous for
diagrams with a larger number of loops. The calculation must be carried
out loop by loop, with more and more terms and flow reversals to keep
track of. Besides, with multiloops there is a huge number
of ways in choosing the independent
loops and their order of integrations, each giving
very different intermediate results though at the end
they must all yield the same contributing poles. It is not known
a priori how
to make the best choice to maximally
simplify the intermediate calculations.

To illustrate these points we shall work out in this
subsection a four-loop example and obtain its contributing poles
in two different ways.

Consider Fig.~2, with the following choice of independent loops:
$a=(4,8,12,13,7)$, $b= (5,9,3,11,12,8) , c=(13,12,10,1,6)$, and
$d=(10,12,11,2)$. Note that lines 8 and 13 are supposed not to 
intersect in the diagram. 
We shall carry out the integrations
in the order $a,b,c,d$ as much as possible.

\begin{figure}
\vskip -0 cm
\centerline{\epsfxsize 3 truein \epsfbox {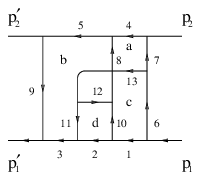}}
\nobreak
\vskip -2 cm\nobreak
\vskip .1 cm
\caption{A four-loop (Feynman) flow diagram.}
\end{figure}

The first integration over loop $a$  yields
$a(4)$ and $a(7)$. 

We do the $b$ integration next. The only lines common to loops $a$
and $b$ are 8 and  12, but since they are antiparallel to 4 and 7, no flow
reversal takes place in carrying out the $b$ integration.
After the $b$-integration we get four terms, which
for brevity shall be written
together as additions: $[a(4)+a(7)][b(3)+b(11)]$.

Line 12 of loop $c$ is also in loop $a$ and loop $b$, and line
13 of loop $c$ is in loop $a$. Since line 12 is antiparallel to 4 and 7
in loop $a$, and antiparallel to 3 and 11 in loop $b$, it suffers
no flow reversal at loop $c$. Similarly line 13, being antiparallel to
lines 4 and 7 in loop $a$, also has no flow reversal. Thus after
the $c$ integration, we get $[a(4)+a(7)][b(3)+b(1)][c(1)+c(10)]$.

The final $d$-integration
is a bit complicated because loop $d$ contains some of these poles
from previously integrations so we are sometimes forced to take
the loop $d.c$ or the loop $d.b$ instead of $d$ itself. The
final result contains 10 terms:
\begin{eqnarray}
&&[a(4)+a(7)]b(3)c(1)[d(10)+d(11)]+\nonumber\\
&&[a(4)+a(7)]b(3)c(10)d.c(2)+\nonumber\\
&&[a(4)+a(7)]b(11)[c(1)d.b(2)+c(10)d.c(2)].
\end{eqnarray}

To summarize, we have obtained ten contributing poles:
(7,3,1,10),
(7,3,1,11), (7,3,10,2), (7,11,2,1), (7,11,2,10), as well as another five
with line 7 replaced by line 4.

Let us now illustrate another way to get the same result, by
choosing this time the four independent loops to be
 $a=(4,8,12,13,7), b=(5,9,3,11,12,8)$,
$e=c.d=(1,6,13,11,2)$, and $d=(10,12,11,2)$, and try to carry
out the integration in the order $a,b,c.d,d$ as much as possible.

These loops are what we shall later call the {\it natural loops}
for the contributing pole (7,3,1,10).
They are obtained first by removing the lines 7,3,1,10 from the
original diagram, and then inserting one of them back 
at a time to get the four loops.

The first two integrations are identical to those before, so we get
$[a(4)+a(7)][b(3)+b(11)]$. Now $e=c.d$ contains the line 11
but not 3, so the next integration involving $b(3)$ gives
$e(1)+e(2)$ but the next integration involving $b(11)$ gives
$c(1)+c(10)$, as $c=d.(c.d)$. The last loop $d$ contains lines
2 and 11, so for some terms the integration over $d$ has to be
changed into integration over $d.b$ or $d.e=c$. The final answer is
\begin{eqnarray}
&&[a(4)+a(7)]b(3)e(1)[d(10)+d(11)]+\nonumber\\
&&b(3)e(2)c(10)+b(11)c(1)d.b(2) +\nonumber\\
&&b(11|3)c(10)d.b(2).
\end{eqnarray}

This results in the same ten contributing poles as before, as it should.

The calculation could be even more complicated if we encounter a line
$j$ which is parallel to a pole line $i$ of an earlier loop,
but the relative magnitude of $q_{j-}$ and $q_{i-}$ can be either way.
In that situation we must divide this flow diagram into two, one in
which $q_{j-}<q_{i-}$ and line $j$ is not reversed, and the other with
$q_{j-}>q_{i-}$ where line $j$ must be reversed.

\section{Path Method for Finding Contributing Poles}

In this section we propose
a simple (path-) method to obtain the contributing poles.  With this
method there is no need to
declare the independent loops and their order of integrations,
so there is no need to keep track of the complicated flow
reversals either. This makes the method most useful in the
presence of a large number of loops.

We begin by choosing a path $P$ in the flow diagram.
By a path we mean a continuous line (no branches, no loops)
running from beginning to end, with all the arrows on it
 pointing in the same
direction. The thin solid lines in Figs.~3 and 4 are examples of such paths.
By adding branches to the path we can construct trees. A class of these
trees, $T[P]$, turns out to be in one-one correspondence with the
contributing poles. The path method of finding contributing poles is
actually a method to construct the trees in $T[P]$.

From an $\ell$-loop diagram one can obtain trees by removing $\ell$ lines.
We shall refer to these removed lines as the {\it missing lines} for the tree.
The set of all trees so obtained with path $P$
as their common backbone will be denoted by $S[P]$.
From $S[P]$ we select a subset $T[P]$ satisfying the
following {\it directional rule}:
when any one of the $\ell$ missing lines is inserted into the tree,
a loop is formed. If the inserted line around this
loop is parallel to the lines along path $P$, this
tree is rejected. If it is anti-parallel, then this tree is
retained to be a member of $T[P]$.

We assert that the missing lines of any tree in $T[P]$
is a contributing pole of the diagram, and there is actually a one-one
correspondence between contributing poles and
individual trees in $T[P]$. This is the essence of
the {\it path method}.

This method does not restrict what path $P$ one chooses,
but the longer the path the fewer the number of
contributing poles,
and the easier the calculations. So in practice we often
choose the longest path we can manage, though this is not a requirement
of the method. In Sec.~VIIC an example will be shown in which computations
based on two different paths are shown for comparison. The reason why
one can get the same result by choosing different paths $P$,
or equivalently different sets of contributing poles, is because
of the freedom to choose poles from either half-plane each time 
we carry out any integration.

We have implicitly assumed in
these discussions that a path $P$ is chosen after we are given a flow
diagram. This is not strictly necessary. We may start from a Feynman
diagram or a nonabelian cut diagram, without arrows attached, and start
drawing a path on it. This can be taken as
the starting point to determine
possible flow diagrams consistent with this path: arrows on the path
must all point in one direction, other arrows must be installed
not to violate the direction rule to obtain contributing poles.

Before proceeding to prove the path method
let us first see how it can be applied to obtain the contributing poles
of Figs.~1 and 2 very simply.

\subsection{Examples}
For Fig.~1 let us choose the path $P$ to be (5347), shown
in Fig.~3 as thin solid lines.
 Then $S[P]=
\{(P,6),(P,1), (P,2)\}$, and $T[P]=\{(P,6),(P,2)\}$.
The tree $(P,1)$ violates the directional rule for
the following reason so it is
not in $T[P]$. When line
6 is inserted into $(P,1)$, it is parallel to $P$ in
the loop (6153), so it has to be rejected.
With this $T[P]$, the contributing poles are the missing lines so they are
(1,2) and (1,6), agreeing with the result obtained previously.

\begin{figure}
\vskip -0 cm
\centerline{\epsfxsize 3 truein \epsfbox {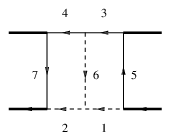}}
\nobreak
\vskip -2.5 cm\nobreak
\vskip .1 cm
\caption{The solid line is the path $P$ used to obtain contributing
poles for Fig.~1.}
\end{figure}

Let us next apply the method to obtain the contributing poles of Fig.~2,
taking $P=(6,13,12,8,5,9)$ as the path (Fig.~4). Then
\begin{eqnarray}
T[P]=\{&&(P,7,2,11),(P,7,2,10), (P,7,1,11),\nonumber\\
 &&(P,7,3,10), (P,7,1,3)\} \ , \nonumber´û
\end{eqnarray}
and five more with 7 replaced
by 4.
The contributing poles are therefore  (7,3,1,10),
(7,3,1,11), (7,3,10,2), (7,11,2,1), and (7,11,2,10),
and another five with 7 replaced by 4, the same 10 terms as before.
The trees in $S[P]/T[P]$ are $\{(P,7,1,2),(P,7,2,3), (P,7,10,11)\}$,
and three more with 7 replaced by 4.
$(P,7,1,2)$ violates the directional rule when the
line 11 is inserted; $(P,7,2,4)$ violates the directional rule when
line 10 is inserted;  and $(P,7,10,11)$ violates the directional rule when
2 is inserted.

\begin{figure}
\vskip -0 cm
\centerline{\epsfxsize 3 truein \epsfbox {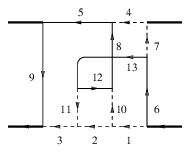}}
\nobreak
\vskip -3 cm\nobreak
\vskip .1 cm
\caption{The solid line is the path $P$ used to obtain the
contributing poles of Fig.~2.}
\end{figure}

\subsection{Proof}
A tree $t\in S[P]$ defines a set of independent loops
${\cal N}[t]$ of the original diagram by filling in the missing lines
one at a time. The special feature
of ${\cal N}[t]$ is that the missing lines are never on the boundary
of two loops. We shall later on refer to these loops as the
{\it natural loops} for the missing lines.

Now we proceed to the proof of the path method. We assume we always
close the integration contour in the half-plane in which poles
reside on lines running in the
opposite direction as those
on $P$.

The proof makes use of the simple fact that the same set of contributing poles
can be computed using any independent loops and any order
of integration.

Removing the pole lines of a contributing pole from the original diagram
gives rise to a tree in $S[P]$. We shall denote the set of all such
tress as $T'[P]$. Our task is to show that $T[P]=T'[P]$.

Take any $t'\in T'[P]$. The removed pole lines clearly satisfy the directional
rule when they are inserted back, because poles are always taken from
those lines running in the opposite direction as $P$. Hence $t'\in T[P]$
and $T'[P]\subset T[P]$.

Conversely, take a $t\in T[P]$, and use the independent loops ${\cal N}[t]$
to compute the contributing poles. The missing lines of $t$ are obviously
one of the pole lines, for according to the directional rule they all
run opposite to the path direction. Hence $t\in T'[P]$ and $T[P]\subset
T'[P]$.

Putting the two together, we get $T[P]=T'[P]$, as desired.

\section{Nonabelian Cut Diagrams}
General methods found in the literature
to compute high energy limits of
Feynman diagrams \cite{POLK,cw} are by and large
valid only in the leading-log approximation (LLA). 
They become virtually powerless if
these leading-log contributions cancel when the
Feynman diagrams are summed,
a situation which unfortunately occurs quite frequently \cite{cw}.
A method was developed recently to bypass this difficulty,
by allowing the cancellations to occur before the high energy
limit is taken. The cancellations are
incorporated into the individual {\it nonabelian cut diagrams} \cite{FHL,LL},
whose spacetime amplitudes (for onshell diagrams)
turn out to differ from the corresponding
Feynman diagram only by having the denominators
$(q_i^2-m^2+i\epsilon)^{-1}$ of certain
propagators replaced by the corresponding
Cutkosky propagators
$-2\pi i\delta(q_i^2-m^2)$. The advantage of the nonabelian
cut diagrams is that the sum of Feynman
diagrams is the same as the sum of nonabelian cut diagrams, 
but in the latter cancellations took place
before the high-energy limit is taken, so
their leading-log contributions (LLA) survive the sum.
For this to happen it is clearly necessary for the LLA of a
nonabelian cut digaram to have a smaller $\ln s$ power than
the corresponding Feynman diagram, if the sum of the LLA
contributions of the latter is to vanish. This is actually
made possible by the presence of the Cutkosky propagators.

For high-energy two-body  ({\it e.g., quark-quark}) scattering,
the Cutkosky propagators occur only on the top quark lines.
In the high energy limit, it can be shown that the
combination $q_i^2-m^2$ is actually proportional to the
`$-$' momentum on that line, so a $\delta$-function is that
variable is a $\delta$-function of the `$-$' momentum \cite{FHL,FL}.
This has the effect of stopping the `$-$' momentum from flowing through
this line, so as far as the flow diagram is concerned we may think
of these lines as being absent. For the rest of the
nonabelian cut diagram
the flows are constructed in exactly the same way as
in a Feynman diagram, and
contributing poles can be located the same way just as well.

As an example, consider
the nonabelian-cut (flow) diagram of Fig.~5, where
the Cutkosky propagator is located at line 8, indicated there by
a vertical bar ($|$).
Hence the {`$-$'} momentum is absent from lines 8, and also from
line 4 by continuity. We may therefore ignore these two lines in the
rest of the discussions.

To obtain the contributing poles from the path method,
we can choose the path to be $P=(1,5,10,6,3)$, then
$T[P]=\{(P,9),(P,7)\}$, giving rise to
the contributing poles (2,7) and (2,9).

\begin{figure}
\vskip -0 cm
\centerline{\epsfxsize 3 truein \epsfbox {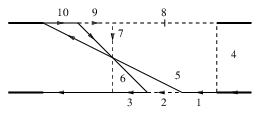}}
\nobreak
\vskip -2.5 cm\nobreak
\vskip .1 cm
\caption{A 3-loop (non-abelian cut) flow diagram. The solid line
represents the path $P$.}
\end{figure}

\section{Dominant Integration Regions in LLA}

Contributing poles, extracted from the path method
or otherwise,
can be used to determine the internal momenta
most important to the loop amplitude. The `+' momenta 
 from the lines of a contributing pole
 $I=(i_1i_2\cdots i_\ell)$ 
are fixed by the pole condition to be
 $q_{i_k+}=(a_{i_k}-i\epsilon)/q_{i_k-}$, and those of any other
line are fixed by momentum conservation. An easy way
to read them out is to use the {\it natural loops}
 discussed
before. These are simply the independent loops containing one and
only one pole line each. 

In LLA a number of simplifications emerge immediately.
For quark-quark scattering in the c.m. system, quark 1
carries a `$-$' momentum $\sqrt{s}$ and quark 2 carries a `$+$'
momentum $\sqrt{s}$. In LLA, where
$|t|$ and squared masses are ignored compared
to $s$, both quarks go straight through by carrying the full forward
momenta with them. In other words, $q_{j-}\simeq
\sqrt{s}$ for every line $j$ of quark 1 (the `bottom lines'), 
and $q_{j+}\simeq\sqrt{s}$
for every line $j$ of quark 2 (the `top lines'). 
This means that we can ignore the 
contributing poles with a pole line on top, for $q_{j+}$
of a top line is $\sqrt{s}$ and  not determined by the pole 
condition above.
In other words, if we insist on taking a pole there, then this term
will not contribute in the LLA.

The two-body amplitude
for a flow diagram, after the `+' integration is
performed,  can be written as
\begin{eqnarray}
M&=&\int\left(\prod_{b=1}^\ell d^2k_{b\perp}\right)F,\nonumber\\
F&=&\int_R\left(\prod_{b=1}^\ell dx_b\right)G,\nonumber\\
G&\equiv&{N\over D}\equiv{N\over\prod_{j=1}^nd_j},\label{ampl}
\end{eqnarray}
where $x_b=k_{b-}/\sqrt{s}$ and $Q_j=q_{j-}/\sqrt{s}$ are the scaled
`$-$' momenta. In practice $x_b$ are chosen from
the $Q_j$'s of types (i) and (ii) below.
The integration region $R$ of $x_b$ is determined by
the $n$ flow-diagram conditions $Q_j\ge 0$. 

The denominator $D=\prod_jd_j$ is derived from the 
denominators of the propagators
$1/d(q_i)$, scaled in some convenient way as follows.
(i) If line $j=i_k$ is part of the contributing pole $I$, then
$d_j$ is defined to be the scaled residue $Q_j$; 
(ii) if line $j$ is a top line, then $d_j\equiv d(q_j)/s
= \pm Q_j-a_j/s+i\epsilon\simeq \pm Q_j+i\epsilon$,
where the sign in front of $Q_j$ is $+/-$ if
the arrow on line $j$ is parallel/antiparallel to the `+' flow
of quark 2; (iii)
for any other line $j$, $d_j$ is equal to $d(q_j)$ evaluated at the
contributing pole, so
\begin{eqnarray}
d_j=Q_j\sum \left(\pm{a_{i_k}\over Q_{i_k}}\right)-a_j,\label{dj}
\end{eqnarray}
where the sum is taken over lines $i_k$ in the same natural loops
as line $j$, with an appropriate sign.

For convenience we will label lines of these three types
by different  indices: index $p$ (for `{\it p}ole') for type (i),
$t$ (for `{\it t}op') for type (ii), and $s$ (for `{\it s}ide')
for type (iii). We shall retain the index $j$ to
denote any of them in general.

The numerator factor $N$ consists of
all the rest, including the 
vertex factors and factors of $\sqrt{s}$ discarded by $D$.

It should be noted that there are no explicit factors of `$i$'
hidden in $M$, except those explicity contained in the vertices
and those appearing as $i\epsilon$ in the propagators. An $\ell$-loop
Feynman diagram has an explicit factor $(-i)^\ell$, and this is
cancelled by the $\ell$ factors of $2\pi i$ from contour integration,
leaving behind no explicit factors of $i$. This observation is important
in determining how the imaginary part of a scattering amplitude arises.

For nonabelian cut diagrams with $c$ cuts, the Feynman propagator
$1/d(q)$ at each cut line is 
replaced by the Cutkosky propagators $-2\pi i\delta(q^2-m^2)$ \cite{FHL,FL}, so an explicit factor $(-i)^c$ will emerge.

From (\ref{ampl}) and the rules for $d_j$,
 it would appear that the integral $F$ diverges
at the boundaries $Q_p=0$ and $Q_t=0$. Actually because of {\it obstructions} from the side lines $s$, the singularity in the $Q_p$
variable is cancelled so
there are no divergences at $Q_p=0$. This is so because as
 $Q_p\to 0$,
 the `+' momentum $q_{p+}\simeq a_p/Q_p$ becomes very large.
At some point it will become much smaller than all 
the $Q_s$, whence $d_s\simeq (Q_s/Q_p)a_s$ for any line $s$
in the natural loop of $p$. This washes out
the factor $Q_p$ in $d_p$, leaving behind no divergence
at this boundary.

A divergence does occur at $Q_t=0$, but this divergence is an artifact
 of our high-energy approximation of dropping $\xi/s\equiv\mp(a_t-i\epsilon)/s$
compared $Q_t$, where $\xi$ is of the order of the squared masses and
the squared momentum transfer $-t$. If we restore it by installing
 a cutoff $\xi/s$ at these boundaries, the divergences will be absent and they 
will be turned into enhancement
factors of $s$. If the enhancement is logarithmic, the value of
$\xi$ does not matter in the LLA, and that will be the case in gauge theories.
But if it is power-like, then
the coefficient of the power dependence would depend
 on  $\xi=\mp(a_t-i\epsilon)$, and its effective value could be determined
only after the transverse-momentum integrations. 

The integral $F$, thus enhanced, receives contributions
in the form
\begin{eqnarray}
F\simeq\int_{(\xi/s)}{dQ'_1\over Q_1^{'m_1}}F_1,\label{f1}
\end{eqnarray}
where $Q'_1$ is either one of the $Q_t$'s, or the radial
variable of several of them that are linearly independent.
As it will become clear shortly this will be the 
smallest of all the `$-$' variables in the dominant
integration region $R_0$.

In the region $Q'_1\ll Q_{p}$, we may set the ratios
$Q'_1/Q_{p}=0$ in all remaining $d_s$. This removes
obstructions from some of the side lines,
so that the integrand of
$F_1$ may now encounter singularity again 
in some variable $Q_2'$, say like $1/Q_2^{'m_2}$, 
with $m_2\ge 1$. This new singular variable 
$Q_2'$ would be equal to some $Q_t$ or $Q_p$, or the 
radial variable of several
of them. Now we have
\begin{eqnarray}
F_1\simeq\int_{B_1Q_1'}{dQ_2'\over Q_2^{'m_2}}F_2,\label{f2}
\end{eqnarray}
for  some $B_1\gg 1$. Similarly, in the region 
$Q_1'\ll Q_2'\ll Q_p$
for the remaining pole lines $p$, $Q_2'/Q_{p}$ can also be set equal
to zero, thus removing further obstructions from even more
side lines. This enables another singular variable
$Q_3'$ to emerge, and so on.
Continue this way until no further
singularities are encountered, we get
\begin{eqnarray}
&&F\simeq\int_{(\xi/ s)}{dQ_1'\over Q_1^{'m_1}}\int_{B_1Q_1'}{dQ_2'\over 
Q_2^{'m_2}}\int\cdots\int_{B_{v-1}Q_{v-1}'}{dQ_v'\over Q_v^{'m_v}}F_{v+1}.
\nonumber\\
&&\label{fv}
\end{eqnarray}
The integrand $F_{v+1}$ is assumed  to be regular
so its $Q_i'$ dependences can all be put equal to zero.
All $B_i\gg 1$.

The dominant region of integration $R_0$ is then given by
\begin{eqnarray}
R_0=\{\xi/s\le Q_1'\ll Q_2'\ll \cdots\ll Q_v'\ll 1\},\label{r0}
\end{eqnarray}
from which we can work out where the `+' momenta are located as well.
The transverse momenta $k_{b\perp}$ are all of the same order as
the momentum transfer $\sqrt{-t}$.

In gauge theories only logarithmic enhancements occur. This means
all $m_i=1$, and 
\begin{eqnarray}
F\simeq {F_{v+1}\over v!}(\ln s)^v.\label{lns}
\end{eqnarray}
 For an $\ell$-loop Feynman diagram, the maximum
enhancement is $\sim(\ln s)^\ell$. For an $\ell$-loop nonabelian
cut diagrams with $c$ cuts, the maximum enhancement is $\sim
(\ln s)^{\ell-c}$. We shall refer to diagrams with these maximal
enhancements as {\it saturated}, and these are the diagrams of
most interestes to us in LLA. Diagrams with less enhancements will be called
{\it unsaturated}. A number of saturated and unsaturated diagrams
are considered in the next section as concrete examples to illustrate the procedures here. For saturated diagrams we will also
work out the coefficient of the leading-log term.

\section{Examples}

\subsection{Scalar Ladder Diagram}

\begin{figure}
\vskip -0 cm
\centerline{\epsfxsize 3 truein \epsfbox {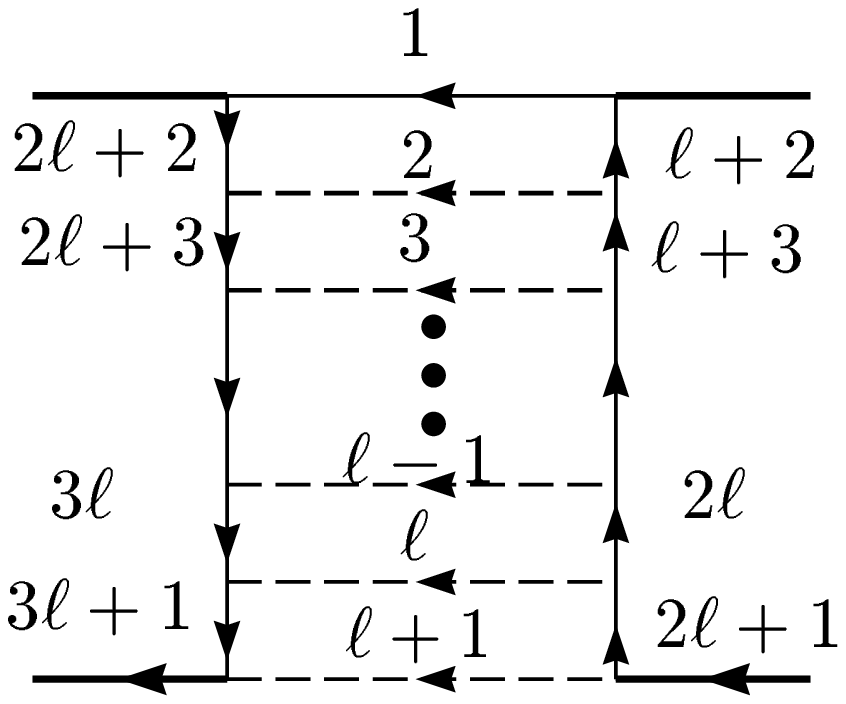}}
\nobreak
\vskip -0cm\nobreak
\vskip .1 cm
\caption{Ladder diagram for scalar quarks and gluons. The path $P$
is indicated by the light solid line and the poles indicated by the dotted
lines. There is only one flow diagram and one contributing pole in this case.}
\end{figure}

Consider the ladder diagram Fig.~6 for scalar quarks and scalar gluons.
There is only one non-zero flow diagram, as shown, and in it
there is only one contributing pole, namely
$I=(2,3,\cdots,\ell+1)$, indicated by the dotted lines. 
The path $P$ from which this contributing pole is obtained
is drawn as a light solid line in the diagram.

In the language of the last section,
the pole lines are $2\le p\le \ell+1$, the top line is
$t=1$, and the side lines are $\ell+2\le s\le 3\ell+1$.
 
The independent `+' momenta at the pole lines are given by
\begin{eqnarray}
q_{p+}\sqrt{s}&=&(a_p-i\epsilon)/Q_p,\label{1+}
\end{eqnarray}
with $Q_{\ell+1}\simeq 1$ in LLA because it is a bottom line.
Thus all these `+' momenta except $q_{\ell+1}$ are capable of
being large if the corresponding $Q_p$ is small enough. The
`+' momenta carried by the side lines can most easily be read off
from the natural loops, which are rectangles
bounded below by the
line $p$ and bounded above by the top line 1.

Following the discussions of last section, 
the top line 1 is the unique candidate 
for the first singular variable $Q_1'$, and indeed it is with
$m_1=1$. In the region $Q_1'\ll Q_j$ for $j>1$, obstructions
from lines $\ell+2$ and $2\ell+2$ are removed, resulting in
$d_{\ell+2}=-a_{\ell+2}$ and $d_{2\ell+2}=-a_{2\ell+2}$. This
allows a new singular
structure to emerge with $Q_2'=Q_2$ and $m_2=1$. This in turn
removes the obstruction from lines $\ell+3$ and $2\ell+3$ in
the region $Q_1'\ll Q_2'\ll Q_j$ for $j>2$, etc. Continuing this way,
we obtain $Q_j'=Q_j$ and $m_j=1$ for $1\le j\le\ell$. Thus the diagram
is saturated, and we obtain the amplitude to be
\begin{eqnarray}
F={1\over\ell!}(\ln s)^\ell\prod_{i=\ell+2}^{3\ell+1}(-a_i).\label{ex1}
\end{eqnarray}
In obtaining this expression, we have set the numerator $N$ of the
integrand to be 1.

The integration region is given by $R_0=\{\xi/s\le Q_1\ll Q_2\ll
\cdots\ll Q_\ell\ll 1\}$. According to (\ref{1+}), the `+'
momenta are strongly ordered in the opposite way because the $a_p$'s
are all of the same order. The virtualities of the side lines $s$
are all spacelike and of order 1, $q_s^2=-a_s=-q_{s\perp}^2$.
In other words, the dominant momenta of the virtual gluons come
from the multi-Regge region, the same as those used in
the dispersion-relation approach \cite{bfkl}.

\subsection{Crossed Ladders}

When the rungs of the ladders are crossed, the scalar diagram will
no longer be saturated. This could be inferred from the example above and 
the $s$-channel dispersion relation, but let us see how to obtain this 
conclusion directly from the Feynman diagram, and how unsaturated
it is. 

\begin{figure}
\vskip -0 cm
\centerline{\epsfxsize 3 truein \epsfbox {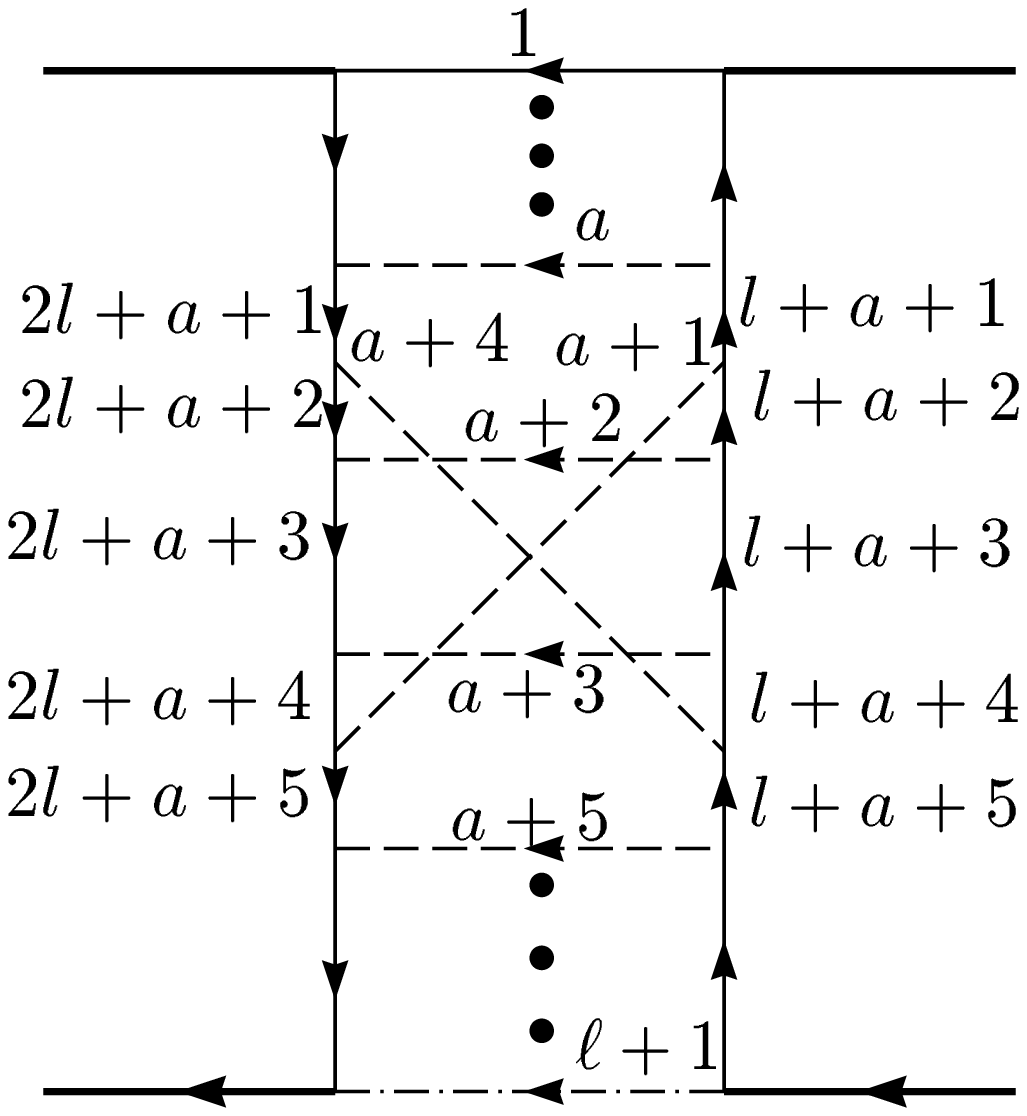}}
\nobreak
\vskip -0 cm\nobreak
\vskip .1 cm
\caption{A crossed ladder diagram, with path $P$ given by the light
solid line and the pole lines given by the dotted lines. The rungs
above $a$ and below $a+5$ are all uncrossed.}
\end{figure}

Consider Fig.~7, which is obtained from Fig.~6 by crossing two 
 rungs separated by $r=2$ horizonatal rungs in between. The path
$P$ and the contributing pole remain unchanged. As before,
we let $Q'_j=Q_j$ for $1\le j\le a$,  and let these $Q'_j$'s
to be strongly ordered as before. Then $m_j=1$
just as in the previous example. The question is what happens
when we come to the region where the rungs are crossed.

Every `+' momentum
$q_{i+}\sqrt{s}\ (1\le i\le 3\ell+1)$ is a linear combination of some $a_p/Q_p\ (2\le p\le \ell+1)$. We shall use the symbol
$[p_1p_2\cdots p_k]$ to represent this `+' momentum 
if it receives contributions
from $p=p_1,p_2,\cdots,p_k$ {\it in the crossed region}. 
Similarly, its `$-$' momenta are linear combinations of
$Q_1$ and $Q_p\ (2\le p\le \ell)$, and those 
from the crossed region that contribute
to the `$-$' momentum of a particular line will be
enclosed between angular brackets $\langle\cdots\rangle$.

The `$-$' and `+' momenta contributions
for the side line $s=\ell+a+k$ on the right ($1\le k\le 5)$
are $\langle a,\cdots,a+k-1\rangle[a+k,\cdots,a+5]$. For the side lines
$s=2\ell+a+k$ on the left, they are $\langle a\rangle[a+4,a+2,a+3,a+1,a+5]$
for $k=1$, $\langle a,a+4\rangle[a+2,a+3,a+1,a+5]$ for $k=2$, $\langle a,a+4,a+2\rangle
[a+3,a+1,a+5]$ for $k=3$, $\langle a,a+4,a+2,a+3\rangle [a+1,a+5]$ for $k=4$,
and finally $\langle a,a+4,a+2,a+3,a+1\rangle [a+5]$ for $k=5$. There
is no way to strongly order the variables
$Q_{a+1},Q_{a+2},Q_{a+3},Q_{a+4}$ in the crossed region
to get rid of all the obstructions. 
Whatever that works on the right hand side will
fail on the left hand side, and vice versa. 
The only way out is to have these four to be of the same order,
for then the ratio of any two of these four would be of order 1, and
all the obstructions from the side lines would disappear.
Their common radial variable $Q'_{a+1}=
(\sum_{i=1}^4Q_{a+i}^2)^{1/2}$ is singular, with $m_{a+1}=1$, because
 $Q_{a+1}^{'3}dQ_{a+1}'/Q_{a+1}^{'4}
=dQ_{a+1}'/Q'_{a+1}$.
From there on, everything looks like the uncrossed ladder again,
so $Q_j'=Q_{j+3}$ for $a+2\le j\le p=\ell-3$. The final
integral $F$  is proportional to $(\ln s)^{\ell-3}$,
hence unsaturated. More generally, the same argument shows that
if there are $r$ uncrossed rungs between the two crossed rungs,
then $F\sim(\ln s)^{\ell-r-1}$.

\subsection{Two-Loop QED Diagram}
Consider now the two-loop diagram Fig.~1 for electron-electron
scattering by exchanging 3 photons.

\begin{figure}
\vskip -0 cm
\centerline{\epsfxsize 3 truein \epsfbox {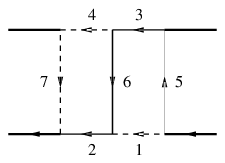}}
\nobreak
\vskip -3 cm\nobreak
\vskip .1 cm
\caption{The solid line is the path $P'$ used to obtain the
contributing poles from Fig.~1, indicated here by dotted
lines. This is a different path than
the one used in Fig.~3.}
\end{figure}

We shall compute this in two ways.
First, using the path $P$ and the contributing poles of Fig.~3, we will
obtain saturated contributions from each of these two contributing poles,
but their sum vanishes so this diagram turns out to be
unsaturated. To see this unsaturation directly, 
we will use another path $P'$
shown in Fig.~8. This path has only one contributing pole
so there can be no chance of a cancellation, and it gives rise to
an unsaturated LLA amplitude. 
In this latter approach we would also be able to compute the coefficient
of the leading log term by LLA calculation if we should want to. 

The numerator $N$ of (\ref{ampl}) in this case comes from the vertices,
and is proportional to $s$. For simplicity we will assume it to be simply
$s$.

The path $P$ from  Fig.~3 gives two contributing poles,
$I_1=(1,2)$ and $I_2=(1,6)$.  
First consider $I_1=(1,2)$. Since both poles lie on the bottom line,
$Q_1\simeq Q_2\simeq 1$, there are no obstructions on the side lines.
Since $Q_3>Q_4$, the integral is
\begin{eqnarray}
F&\simeq&-{s\over a_5a_6a_7}\int_{(\xi/s}{dQ_4\over Q_4}
\int_{Q_4}{dQ_3\over Q_3}\nonumber\\
&\simeq&{s\over (-)^3a_5a_6a_7}\int_{\xi/s}{dQ_4\over Q_4}\int_{Q_4}{dQ_3\over Q_3}
\simeq -{s\over 2 a_5a_6a_7}(\ln s)^2.\nonumber\\
&&\label{2loop1}
\end{eqnarray}
Next consider $I_2=(1,6)$. The pole on 6 causes an obstruction from lines
2 and 7. By choosing $Q_1'=Q_4\ll Q_2'=Q_6$, the obstruction from line 7
is removed but the obstruction from line 2 remains because $Q_2\simeq 1$.
However, since $Q_3\simeq Q_6=Q_2'$, the contribution from $I_2$ is
\begin{eqnarray}
F&\simeq&{s\over (-)^2a_5a_7}\int_{\xi/s}{dQ_1'\over dQ_1'}\int_{B_1Q_1'}{dQ_2'
\over Q_2^{'2}(a_6/Q_2')}\nonumber\\
&\simeq& +{s\over 2a_5a_6a_7}(\ln s)^2.\label{2loop2}
\end{eqnarray}
The sum of the contributions from $I_1$ and $I_2$ vanishes in order
$(\ln s)^2$ so the diagram is unsaturated.

To see this unsaturation directly, choose another path
$P'$ as shown in Fig.~8. The contributing pole is now $I'=(1,7)$.
The obstruction induced by line 7 on lines 2 and 6 block out the factor
$d_4d_7=Q_4Q_7$, so to get a singular integrand for $F$ we must
enlist the help of $Q_3$. If $Q_1'$ is the radial variable of $Q_7$
and $Q_3$, then the integrand of $F$ is proportional to
$Q_1'dQ_1'/Q_3Q_4Q_7d_2\sim dQ_1'/Q_1'$, so the leading contribution
to this diagram is of the order $\ln s$.

\subsection{Four-Loop Diagram}

If the path $P$ for the four-loop diagram Fig.~2 is chosen as in Fig.~4,
then as we have seen there are 10 contributing poles.
For illustration we will look at the contribution from a single one,
$(7,3,10,2)$. We shall see that there will be no 
$\ln s$ enhancement if the diagram is scalar, 
but if it is a QCD diagram then there
will be a linear $\ln s$ enhancement from this contributing pole.

There are two top lines in this diagram, lines 4 and 5.
Since $Q_5>Q_4=Q_7$, the only single-variable candidate for $Q_1'$
is $Q_4=Q_7$. However, the pole in 7 produces an obstruction on
all the other lines in its natural loop: lines 8, 12, and 13. 
With three obstructing lines and only two singular factors, the resulting
$Q_1'$ dependence cannot be singular for a scalar diagram. One could go
on and try to find a singular $Q_1'$ among the radial variables of several
$Q_j$'s, and one would not succeed either. Consequently as 
a scalar diagram it has no $\ln s$ enhancement.

As a QCD diagram we must incorporate the vertex factors into the numerator
$N$ of the integrand of $F$. The vertex factor for a gluon connected
to the top line is $2p_2$, and to the bottom line is $2p_1$.
There are however also three triple-gluon vertices, at the junctions
of lines $(6,7,13)=A, (11,12,13)=B$, and $(8,10,12)=C$.
Each of them contains three terms, but one of the three terms of each
is dotted into $2p_1$ and therefore produce an appropriate combinations
of $q_{i+}$: $g_{7,13}(q_{7+}-q_{13+})$ for $A$, $g_{12,13}
(q_{12+}+q_{13+})$ for $B$, and $g_{8,12}(q_{8+}+q_{12+})$ for
$C$. Since every line in the natural loop of 7 contains $\pm q_{7+}$,
and hence a factor $1/Q_1'$, these three vertex factors can 
make the $Q_1'$ variable much more singular. However, we may use only
two out of the three, for otherwise the $g_{\alpha\beta}$ factors
will lead to a dot product of the (7,4) vertex and the (4,5,8)
vertex, thus producing an extra factor $2p_2\!\cdot\!2p_2=0$. With the help
of two triple-gluon vertices, we get $m_1=1$ and a $\ln s$ enhancement
from the $Q_1'$ variable.

The remaining singular factors for the integrand come- from lines 5
and 10. Since $Q_5>Q_{10}$, if $Q_2'$ comes from a single variable
$Q_j$ we must have $j=10$. This pole at 10 may produce obstructions on
lines of its natural loop (10,12,13,6,1). Those on lines 12
and 13 have already been removed by $Q_1'$, so this leaves obstructions
from lines 1 and 6. The one on 1 is particularly troublesome because
it is a bottom line, so $Q_1\simeq 1$ and the obstruction can never
be removed. For that reason $Q_{10}$ is not a singular variable, and
it can be checked that the radial variable of $Q_{10}$ and one or two
other $Q_j$'s cannot be a singular variable either. The enhancement of
the QCD diagram is therefore just $\ln s$. 

\subsection{Nonabelian Cut Diagram}

As a last example we consider the nonabelian cut diagram, Fig.~5,
treated as a scalar diagram with numerator factor $N=1$.
The path is $P=(1,5,10,6,3)$ and the contributing poles are $(2,7)$
and $(2,9)$. Since in the LLA we would never have to consider any
contributing pole on the top line, we can drop (2,9) and consider
only (2,7).

In either case there is actually a hidden contributing pole at line 1.
This does not show up explicitly in the path method because
the cut line 8 reduces the flow diagram into a two-loop diagram,
hence only two of the three poles show up explicitly. In any case,
since $Q_4=Q_8=0$, we have $Q_1=1$, so the pole at 1 produces
$q_{1+}=a_1$. This together with the $q_{p+}$ obtained from the other
two pole lines uniquely determine all the `+' momenta of all the lines.
However, the contribution from $q_{1+}$ is finite,  it will never 
lead to an obstruction, so in some sense we can just forget about it.

Of the two uncut top lines, $Q_{10}>Q_9=Q_7$, so if $Q_1'$
is given by a single $Q_j$, it would have to be $j=9$. The second singular
variable is $Q_2'=Q_6=Q_{10}\gg Q_1'$, and the integral is
\begin{eqnarray}
F=&&\int_{\xi/s}{dQ_1'\over Q_1^{'2}(-a_7/Q_1')(-a_6)}
\int_{m_1Q_1'}{dQ_2'\over Q_2'(-a_5)(-a_4)}\nonumber\\
&&\int dQ_8(-2\pi i)\delta(Q_8)
\simeq -{\pi i\over a_4a_5a_6a_7}(\ln s)^2.
\end{eqnarray}

\section{Acknowledgement}
This research was supported in part by the Natural Science and Engineering
Research Council of Canada and by the Qu\'{e}bec Department of Education.
Y.J.F. acknowledges the support of the Carl Reinhardt Major Foundation,
and C.S.L. wants to thank Hung Cheng for a stimulating discussion.



\begin{thebibliography}{99}
\bibitem[\dag]{YJF} Electronic address: feng@physics.mcgill.ca
\bibitem[*]{CSL} Electronic address: lam@physics.mcgill.ca
\bibitem{lam} C.S. Lam, hep-ph/9804463.
\bibitem{fadlip} For example, see
V.S. Fadin and L.N. Lipatov, {\it Nucl. Phys.} {\bf B477}
(1996) 767; hep-ph/9802290.
\bibitem{POLK}R.J. Eden, P.V. Landshoff,
D.I. Olive, and J.C. Polkinghorne, {\it The Analytic S-Matrix},
 (Cambridge University Press, 1966).
\bibitem{cw} H. Cheng and T.T. Wu, {\it
`Expanding Protons: Scattering at High Energies'}, (M.I.T. Press, 1987).
\bibitem{bfkl}
L.N. Lipatov,  Yad.~Fiz. 23 (1976) 642 [Sov.~J.~Nucl.~Phys.
23 (1976) 338]; Ya. Ya. Balitskii and L.N. Lipatov, Yad.~Fiz. 28 (1978) 1597
[Sov.~J.~Nucl.~Phys. 28 (1978) 822];
E.A. Kuraev, L.N. Lipatov, and V.S. Fadin, Zh.~Eksp.~Teor.~Fiz.
71 (1976) 840 [Sov.~Phys. JETP 44 (1976) 443]; {\it ibid.} 72 (1977) 377 [{\it
ibid.} 45 (1977) 199];
V. Del Duca, hep-ph/9503226.
\bibitem{bartels} J. Bartels, {\it Nucl. Phys.} {\bf B151} (1979) 293;
{\bf B175} (1980) 365.
\bibitem{hera} H1 Collaboration, hep-ex 9708016;
ZEUS Collaboration, hep-ex 9709021.
\bibitem{reggeon} J. Bartels, {\it Z. Phys.} {\bf C60} (1993) 471;
J. Bartels and M. Wuesthoff, {\it Z. Phys.} {\bf C66} (1995) 157;
M.A. Braun, {\it Nucl. Phys.} {\bf B175} (1980) 365; {\it hep-ph}/9706373.
\bibitem{lam2} See, for example, the references cited in Ref.~\cite{lam}.
\bibitem{FHL}Y.J. Feng, O. Hamidi-Ravari, and C.S. Lam,
{\it Phys. Rev. D} {\bf 54} (1996) 3114.
\bibitem{FL}Y.J. Feng and C.S. Lam, {\it Phys. Rev. D}
{\bf 55} (1997) 4016.
\bibitem{LL}C.S. Lam and K.F. Liu, {\it Nucl. Phys.} {\bf B483} 
(1997) 514; {\it Phys. Rev. Lett.} {\bf 79} (1997) 597. 
\bibitem{verlinde} H. Verlinde and E. Verlinde, hep-th 9302104.
\bibitem{lipatov} L.N. Lipatov, {\it Nucl. Phys.} {\bf B365} (1991) 614; 
R. Kirschner, L.N. Lipatov, and L. Szymanowski, 
{\it Nucl. Phys.} {\bf B425} (1994) 579; {\it Phys. Rev. D} {\bf 51}
(1995) 838.
\end{thebibliography}
\end{document}